\documentstyle[aps,floats,epsf,twocolumn]{revtex}
\newcommand{\beq}{\begin{equation}}
\newcommand{\eeq}{\end{equation}}

\begin{document} \draft

\title{Low Temperature Behavior of the Vortex Lattice in Unconventional
Superconductors}

\author{M. H. S. Amin$^{1}$, Ian Affleck$^{1,2}$ and M. Franz$^3$}
\address{$^1$Department of Physics and Astronomy
and $^2$Canadian Institute for Advanced
Research,\\ University of British Columbia, Vancouver, BC, V6T 1Z1, Canada\\
$^3$Department of Physics and Astronomy, Johns Hopkins University,
Baltimore, MD 21218
\\{\rm(\today)}}\maketitle

\begin{abstract}
We study the effect of the superconducting gap nodes 
on the vortex lattice properties of high temperature superconductors
at very low temperatures. The nonlinear, nonlocal
and nonanalytic nature of this effect is shown to have measurable 
consequences for the vortex lattice geometry and 
the effective penetration depth in the mixed state as measured by 
muon-spin-rotation experiments.
\end{abstract}

\narrowtext

\section{Introduction}

The presence of nodes in the superconducting gap is probably one of the most
significant features of high-$T_c$ superconductors which has attracted 
considerable theoretical and experimental attention in recent years. 
Many experiments\cite{hardy,van} have confirmed the 
existence of gap nodes. These experiments most commonly indicate
an order parameter with $d_{x^2-y^2}$ symmetry\cite{van}.
The fourfold symmetric nature of the $d$-wave order
parameter together with the presence of gap nodes on the Fermi surface
open possibilities for novel effects to be observable in cuprates. 
An early theoretical 
investigation of the weak-field response of a $d_{x^2-y^2}$ superconductor
by Yip and Sauls\cite{ys} predicted a direction dependent non-linear
Meissner effect, associated with the quasiclassical shift of the excitation
spectrum due to the superflow created by the screening currents. Maeda 
{\em et al.} \cite{maeda} reported experimental evidence for such an effect
in Bi$_2$Sr$_2$CaCu$_2$O$_y$, but subsequent experiments failed to confirm 
their findings and the situation remains controversial.
A similar effect was also studied independently by Volovik\cite{volovik}. 
In the mixed state, he predicted a 
contribution to the residual density of states (DOS) proportional to the
inter-vortex distance $\sim\sqrt{H}$. Such contribution was identified in 
the specific heat measurements on YBa$_2$Cu$_3$O$_{7-\delta}$ 
(YBCO) by Moler {\em et al.}\cite{moler}
but later this interpretation was disputed by Ramirez 
\cite{ramirez} who found evidence for similar effect in a conventional 
superconductor V$_3$Si and by others\cite{others}.
Kosztin and Leggett \cite{leggett} predicted
that the nonlocal response at very low temperatures will lead to
a $T^2$ dependence of the penetration depth in clean samples 
in contrast to the linear $T$-dependence obtained from the local theory for 
$d$-wave materials. 

In the mixed state also, it is conceivable that these effects still 
present themselves in some measurable properties such as vortex lattice
geometry, magnetic field distribution etc.. Neutron scattering\cite{keimer}
and STM\cite{maggio} experiments on the high-$T_c$ compound 
YBCO revealed a vortex lattice structure
different from hexagonal -  which is expected for an isotropic superconductor.  However, this may be explainable by penetration depth anisotropy and twin-boundary pinning without involving any effects associated with gap 
nodes\cite{walker}.
Muon-spin-rotation
($\mu$SR) experiments\cite{sonier-thesis,sonier,sonier1,sonier2,sonier3}, 
on the other hand, have demonstrated an
unusual magnetic field dependence in their line-shapes for the magnetic field 
distribution. This has been attributed to a field dependent penetration depth 
- which is expected in the Meissner state because of quasiparticle generation 
at gap nodes\cite{ys}.  It was also modeled using an approach based on the 
Bogoliubov- de Gennes (BdG) equations in a square lattice 
tight-binding model \cite{wang}.

At intermediate fields $H_{c1}<H\ll H_{c2}$ properties of the flux
lattice are determined primarily by the superfluid response of the 
condensate, i.e. by the relation between the supercurrent $\vec j$ and
the superfluid velocity $\vec v_s$. In conventional isotropic 
strongly type-II
superconductors this relation is to a good approximation that of simple 
proportionality,
\beq
\vec j=-e\rho_s\vec v_s,
\label{j0}
\eeq
where $\rho_s$ is a superfluid density. More generally, however, this 
relation can be both {\em nonlocal} and {\em nonlinear}. 
The concept of nonlocal response
dates back to the ideas of Pippard\cite{tinkham} and is related to the 
fact that the current response must be averaged over the finite size of the 
Cooper pair given by the coherence length $\xi_0$. In strongly type-II
materials the magnetic field varies on length scale given by the
London penetration depth $\lambda_0$, which is much larger than $\xi_0$ and
therefore nonlocality is typically unimportant unless there exist strong
anisotropies in the electronic 
band structure\cite{kogan}. Nonlinear corrections arise
from the change of quasiparticle population due to  
superflow which, to leading order,  modifies  the excitation spectrum 
by a quasiclassical shift\cite{tinkham}
\beq
{\cal E}_k=E_k +{\vec v_f}\cdot{\vec v_s},
\label{spec}
\eeq
where $E_k=\sqrt{\epsilon_k^2+\Delta_k^2}$ is the BCS energy.
Again, in clean, fully gapped conventional superconductors this 
effect is typically negligible except when the current approaches the pair
breaking value. In the mixed state this happens only in the close vicinity 
of the vortex cores which occupy a small fraction of the total sample
volume at fields well below $H_{c2}$. The situation changes dramatically
when the order parameter has nodes, such as in $d_{x^2-y^2}$ superconductors.
Nonlocal corrections to (\ref{j0}) become important for the response of 
electrons with momenta on the Fermi surface close to the gap nodes even 
for strongly type-II materials. This can be understood by realizing that
the coherence length, being inversely proportional to the gap\cite{tinkham},
becomes very large close to the node and formally diverges at the nodal 
point. Thus, quite generally, there exists a locus of points on the Fermi 
surface where $\xi \gg\lambda_0$ and the response becomes highly nonlocal.
This effect has been first discussed by Kosztin and Leggett\cite{leggett}
for the Meissner state and by us\cite{franz} in the mixed state. Similarly,
the nonlinear corrections become important in a d-wave superconductor. 
Eq.\ (\ref{spec}) indicates that finite areas of gapless excitations appear
near the node for arbitrarily small $v_s$. This effect has been studied
in the Meissner state\cite{ys} but its consequences have remained largely 
unexplored for the mixed state. 

In our previous works\cite{affleck,franz}, we discussed the effect of
fourfold anisotropies associated with {\em nonlocal} response
on the field and temperature dependence of vortex lattice structure.
In Ref.\cite{affleck}, starting from a phenomenological Ginzburg-Landau
(GL) theory, we derived the leading
fourfold anisotropic corrections to the London equation making the usual assumption that the free energy was an analytic functional of the order parameter and field.
We showed that such corrections result in a field-driven continuous transition
from triangular to square vortex lattice. 
Being derived from a GL theory this London equation is expected to 
lose its validity at low temperatures and a microscopic theory is needed
to address this regime. 
In Ref.\cite{franz}, we investigated the effect of the nonlocality due
to the presence of the nodes in the superconducting gap using a simple
weak coupling microscopic model. At high 
temperatures a nonlocal correction similar to the one suggested in
Ref.\cite{affleck} was obtained.
At low temperatures however, a novel singular behavior was found
directly related to the nodal structure of the gap which 
completely changes the form of the London equation. This singular behavior
has profound implications for the structure of the vortex lattice
which, as a function of decreasing temperature, undergoes a series
of sharp structural transitions and attains a universal limit at $T=0$. 

In Ref.\cite{franz},
we neglected the effect of the {\em nonlinear} terms resulting from
the excitation of quasiparticles at the gap nodes, assuming that they 
are small compared to the nonlocal corrections. In this article, we 
consider both the nonlinear and the nonlocal terms
assuming that the effects are additive and do not affect each other. 
We show that, as we claimed in Ref.\cite{franz}, 
the dominant effect which determines the 
vortex lattice geometry and the effective penetration depth as defined
in $\mu$SR experiments is the nonlocal corrections, while the 
nonlinear corrections play a secondary role at low $T$. 

\section{Generalized London Equation}
\subsection{Nonlinear Corrections}

Let us for the moment neglect any nonlocality effect and focus
on the quasiparticles generated at the gap nodes.
Excitation of quasiparticles at the gap nodes produces a current density
flowing in the  
direction opposite to the superfluid velocity -  sometimes called back-flow. 
The total current can then be written as
\beq
\vec j=-e\rho_s\vec v_s + \vec j_{qp}
\label{j}
\eeq
where $\rho_s$ is the superfluid density and $\vec v_s$ is the 
superfluid velocity defined by\cite{note1}
\beq
\vec v_s={1\over 2}({2 e\over c}\vec A - \nabla \phi) .
\label{vs}
\eeq
$\vec A$ is the vector potential and $\phi$ is the phase of the order 
parameter. The contribution of the quasiparticles generated at the
nodes to the total current is given by\cite{ys}
\begin{eqnarray}
\vec j_{qp}=-4eN_f\int_{FS} ds \vec v_f(s) \int_0^\infty &d\xi&
f(\sqrt{\xi^2+\Delta(s)^2}\nonumber\\ &+& \vec v_f(s)\cdot\vec v_s)
\label{j1}
\end{eqnarray}
where $N_f$ is the density of states at the Fermi surface and $s$ is a 
parameter which represents a point on the Fermi surface. 
$\Delta(s)$ is the superconducting
gap which in general can have $s$-wave, $d$-wave, or other symmetries. 
At zero temperature Eq.(\ref{j1}) leads to
\beq
\vec j_{qp}=-4eN_f\int ds \vec v_f(s) \theta (-\vec v_f\cdot\vec v_s
-|\Delta|)\sqrt{(\vec v_f\cdot\vec v_s)^2-\Delta^2}
\label{j2}
\eeq

\begin{figure}
\epsfysize 7.5cm
\epsfbox[20 220 500 700]{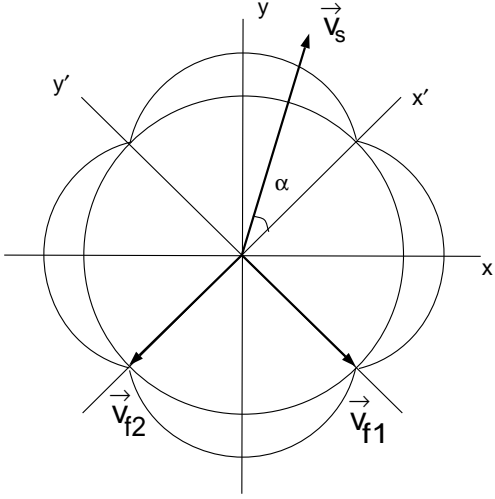}
\caption[]{Circular Fermi surface with a $d_{x^2-y^2}$ gap. Quasiparticles
will be excited at nodes marked by $\vec v_{f1}$ and $\vec v_{f2}$ opposite to
$\vec v_s$.}
\label{fig1}
\end{figure}

At higher temperatures however, this gives the first term in the 
Sommerfeld expansion which is a good approximation as long as 
$T<T^\star=T_c(H/H_0)$, where $H_0$ is of order of the
thermodynamic critical field, $H_c$ \cite{ys}. The presence of
the $\theta$-function in Eq.(\ref{j2}), results in excitations
only at the nodes in the opposite direction to $\vec v_s$.
Fig.\ref{fig1} illustrates a circular Fermi surface with a $d$-wave
gap. The quasiparticles are excited at the nodes marked by
$\vec v_{f1}$  and  $\vec v_{f2}$ in the opposite direction to $\vec v_s$.
For a small enough $v_s$, the excitations stay very close to the
gap nodes. Therefore, one can linearize the gap function near the nodes
writing $\Delta(\theta)\simeq \gamma \Delta_0 \theta$
with $\Delta_0$  the maximum gap and $\gamma$ defined by
\beq
\gamma = {1\over \Delta_0}[{d\over d\theta}\Delta(\theta)]_{\rm node}
\label{gamma}
\eeq
The component of $\vec j_{qp}$ along the $x^\prime$-direction which is
diagonal to the $x$ and $y$ ($a$ and $b$) directions 
(as illustrated in Fig.\ref{fig1}) is then 
\begin{eqnarray}
j_{qpx^\prime}&=&4eN_fv_f\int_{-\theta_c}^{\theta_c}{d\theta \over 2\pi}
\sqrt{(v_fv_s\cos\alpha)^2-|\gamma\Delta_0\theta|^2}\nonumber\\
&=&e\rho_s{v_s^2 \over v_0} \cos\alpha |\cos\alpha |=e\rho_s{v_{sx^\prime}
|v_{sx^\prime}|\over v_0}
\label{jx}
\end{eqnarray}
$\alpha$ is the angle between $v_s$ and $x^\prime$-axis, 
$\rho_s=N_f v_f^2$ is the superfluid
density,  $v_0=\gamma\Delta_0/v_f$ is some characteristic
velocity and $\theta_c$ is a cut off imposed by the $\theta$-function
in Eq.(\ref{j2}). Similarly the $y^\prime$-component is
\beq
j_{qpy^\prime}=e\rho_s{v_{sy^\prime}|v_{sy^\prime}|\over v_0}
\label{jy}
\eeq
and the total current thereby becomes
\beq
\vec j=-e\rho_s[\vec v_s - (v_{sx^\prime}|v_{sx^\prime}|\hat x^\prime
+v_{sy^\prime}|v_{sy^\prime}|\hat y^\prime)/v_0].
\label{js2}
\eeq
The nonanalytic nature of the effect is evident from this equation.  

It is now possible to write a free energy in such
a way  that Eq.(\ref{js2}) can be
obtained by minimization with respect to $\vec A$.
Keeping in mind that$\vec j=(c/4\pi)\nabla\times\vec B =
(c/4\pi)\nabla\times\nabla\times\vec A$ and $\partial/\partial \vec v_s
={c\over e}\partial/\partial \vec A$, the corresponding free energy 
density can be written as
\beq
f=\rho_s\left[{1\over 2}v_s^2-{1\over 3v_0}(|v_{sx^\prime}|^3+
|v_{sy^\prime}|^3)\right]+{B^2\over 8\pi}
\label{f}
\eeq

In general, it is possible to solve Eq.(\ref{js2}) for $\vec v_s$ in terms 
of $\vec j=(c/4\pi)\nabla\times\vec B$, substitute it into Eq.(\ref{f})
 and write a London free energy 
only in terms  of $\vec B$ and its derivatives. However, instead of
solving Eq.(\ref{js2}) exactly, we find $\vec v_s$ perturbatively assuming
that the nonlinear part is much smaller than the linear part. This
results in a polynomial correction to the London equation which is more 
convenient for the numerical purposes. To first order in perturbation theory:
\begin{eqnarray}
\vec v_s={c\over 4\pi e \rho_s}&[&\nabla\times\vec B 
+ {c\over 4\pi e \rho_s v_0}
((\nabla\times\vec B)_{x^\prime}|\nabla\times\vec B|_{x^\prime}\hat x^\prime
\nonumber\\ &+&(\nabla\times\vec B)_{y^\prime}|\nabla\times\vec B|_{y^\prime}
\hat y^\prime))]
\end{eqnarray}
Substituting this into Eq.(\ref{f}) and keeping the lowest order terms, 
the London free energy density becomes
\begin{eqnarray}
f_L={1\over 8\pi}\Bigl[
B^2&+&\lambda_0^2(\nabla\times\vec B)^2+({2\pi\over 3\gamma})
{\xi_0\lambda_0^2\over B_0}(|(\nabla\times\vec B)_{x^\prime}|^3
\nonumber\\&+&|(\nabla\times\vec B)_{y^\prime}|^3)\Bigr]
\label{fL1}
\end{eqnarray}
where $\lambda_0=\sqrt{c^2/4\pi e^2\rho_s}$ is the zeroth order 
penetration depth, $\xi_0=v_f/\pi\Delta_0$
is the coherence length, $B_0\equiv \phi_0/2\pi\lambda_0^2$ is a characteristic
field of the order of $H_{c1}$ and $\phi_0=\pi c/e$ is the flux quantum. 
The first two terms in Eq.(\ref{fL1}) are the ordinary London
free energy terms while the remaining two terms constitute the leading
nonlinear correction. 
 For magnetic fields in the $z$-direction, 
the London free energy density becomes
\beq
f_L={1\over 8\pi}\Bigl[B^2+\lambda_0^2(\nabla B)^2+({2\pi\over 3\gamma})
{\xi_0\lambda_0^2\over B_0}(|\partial_{x^\prime}B|^3+|\partial_{y^\prime}B|^3)
\Bigr]
\label{fL2}
\eeq
and the corresponding London equation will be 
\beq
-\lambda_0^2\nabla^2B+B-({2\pi\over \gamma}){\xi_0\lambda_0^2\over B_0}(\partial_
{x^\prime}^2B
|\partial_{x^\prime}B|+\partial_{y^\prime}^2B|\partial_{y^\prime}B|)=0
\label{London}
\eeq
A similar London equation is also derived by Zutic and Valls
who investigated the effect in the Meissner state\cite{valls}.

The most commonly used form for a d-wave gap is $\Delta(\theta)=\Delta_0
\cos (2\theta)$. In this case, Eq.(\ref{gamma}) leads to a $\gamma=2$. 
As discussed in more detail below, in order
to find the magnetic field distribution 
in a vortex lattice, one has to insert a source term
$\Sigma_j \rho(\vec r - \vec r_j)$ on the right hand side of Eq.(\ref{London}) 
with $\vec r_j$ being the position of the vortices in the lattice.
The function $\rho(\vec r)$ takes into account the vanishing of the 
order parameter at the center of the vortex cores. 
Numerically, it is more convenient to work in  Fourier space rather than
in real space. However, because of the nonanalyticity of the nonlinear
term, the Fourier transformation of this term cannot be done by 
simple convolution integrals and
numerical techniques such as Fast Fourier Transformation (FFT) 
are required.
Fourier transforming Eq.(\ref{London}) with a proper source term on the right
hand side yields
\beq
B_{\vec k}+\lambda_0^2 k^2 B_{\vec k}-G_{\rm nl}(\vec k,B_{\vec k})=\bar B F(\vec k)
\label{ldn}
\eeq
where $\vec k$ is a reciprocal lattice wave vector,  $G_{\rm nl}$ is the
Fourier transform of the nonlinear term and $\bar B$ is the average
magnetic field. The cut off function $F(\vec k)$ comes from the Fourier 
transformation of the source term and removes the divergences
by cutting off the momentum sums.

\subsection{Nonlocal Corrections}

One can add the nonlocal effect to Eq.(\ref{ldn}) by replacing the second
term with a nonlocal one as discussed in Ref.\cite{franz}. The resulting
London equation will be
\beq 
B_{\vec k}+{\cal L}_{ij}(\vec k)k_ik_j B_{\vec k}-G_{\rm nl}(\vec k,B_{\vec k})
=\bar B F(\vec k). 
\label{ldn2}
\eeq 
Sums over $i$ and $j$ are implicit here. ${\cal L}_{ij}(\vec k)$ is related to
the electromagnetic response tensor $\hat Q(\vec k)$ defined in 
Ref.\cite{franz} by
\beq
{\cal L}_{ij}(\vec k)=Q_{ij}(\vec k)/\det \hat Q(\vec k).
\label{L}
\eeq
The electromagnetic response tensor describes the linear response of 
a superconductor to an applied magnetic field. It has a particularly simple
form in the gauge where $\nabla\phi =0$ and $\vec v_s$ is proportional to the
vector potential $\vec A$. At $T=0$ one then obtains\cite{franz},
\begin{equation}
Q_{ij}({\vec k})={1\over \lambda_0^2}\int ds
\hat v_{fi}\hat v_{fj}
{2{\rm arcsinh\ }y \over y\sqrt{1+y^2}},
\label{Q-nono}
\end{equation}
where $y=\vec v_s\cdot\vec k/2\Delta(s)$. The expression for $\hat Q(\vec k)$
becomes more complicated in an arbitrary gauge and was discussed by 
Millis\cite{millis}. In the present context, however, it is easy to verify 
that, because of the particular way $\hat Q(\vec k)$ enters the London equation
(\ref{ldn2}), this is gauge invariant as required for an equation containing 
an observable quantity. 

In the local case we have 
${\cal L}_{ij}(\vec k)=\lambda_0^2\delta_{ij}$
which leads back to Eq.(\ref{ldn}). Taking $G_{\rm nl}$ to the
right hand side of Eq.(\ref{ldn2}), $B_{\vec k}$ can be obtained by
\beq
B_{\vec k}={\bar B F(\vec k) + G_{\rm nl}(\vec k,B_{\vec k}) \over
1+{\cal L}_{ij}(\vec k)k_ik_j}.
\label{bk}
\eeq
We use this equation to find $B_{\vec k}$ iteratively for a specific 
lattice geometry. Having $B_{\vec k}$, the free energy can be 
easily calculated using
\beq
{\cal F}={\cal F}_{\rm nl} + \sum_{\vec k}[1+{\cal L}_{ij}(\vec k)k_ik_j]B_{\vec k}^2 
\label{fe}
\eeq
where ${\cal F}_{\rm nl}$ is the free energy due to the nonlinear parts as in 
Eq.(\ref{fL2}).

\subsection{Vortex Source}

The functional form of $\rho(\vec r)$ and thereby $F(\vec k)$
depends on the detailed form of the order parameter at the center of the vortex
and therefore requires a more fundamental theory to be evaluated.
In the Ginzburg-Landau (GL) limit, there have been some conjectures about the
profile of the order parameter at the vortex center and the form of the
source term \cite{brandt,clem,hao,yaouanc}. 
Brandt\cite{brandt} suggests a Gaussian from for the source term
\beq
\rho(\vec r)=(\phi_0/2\pi\xi_0^2)e^{-r^2/2\xi_0^2}
\label{source}
\eeq
which leads to
\beq
F(k)=e^{-\xi^2 k^2/2}.
\label{cutoff}
\eeq
Clem \cite{clem} on the other hand, assumes the order parameter to vanish
as $r/\sqrt{r^2 + \xi_v^2}$ at the center of the vortex core near $T_c$, 
where $\xi_v$ is a variational parameter proportional to $\xi_0$.
It is not difficult to show that this form of the order parameter leads to a
squared Lorentzian source term \cite{amin}.
\beq
\rho(r)={\Phi_0 \over \pi}{\xi_v^2 \over (\xi_v^2 + r^2)^2},
\hspace{.3cm} F(k)=u K_1(u)
\eeq
where $K_1$ is the modified Bessel function. In the extreme type-II 
case, $\xi_v=\sqrt{2}\xi_0$ and $u=\sqrt{2}\xi_0 k$.
This form of the source term
shows quite a good agreement with the exact solution of the GL equation
in the vortex state\cite{brandt2}. 
At low temperatures however, the GL equation is  
not applicable. The most commonly assumed form for the order parameter near
the vortex core which gives a good fit to the numerical solutions of
the BdG equation \cite{gigy} is
$\psi(r) \propto \tanh(r/\xi)$.
The cut off function resulting from this form of the order parameter
fits quite well with the Gaussian form of Eq.(\ref{cutoff}) 
up to high Fourier modes\cite{amin}. It is worth pointing out that none 
of the models for the source terms discussed above,
consider the $d$-wave nature of the order parameter or other anisotropies 
which might be important for the vortex lattice properties. 
Recent numerical computations within the self-consistent BdG theory\cite{ft}
for a $d$-wave vortex indicate an order parameter with a relatively 
weak four-fold anisotropy and an amplitude relaxing to its bulk 
value as $\sim 1/r^2$ far from the core even at $T=0$.  
Therefore, a more
careful consideration is required to formulate a reasonable model for the 
vortex core in the London theory. We leave a
detailed discussion of the vortex core to a separate paper\cite{amin}
and from now on assume the Gaussian form of Eq.(\ref{source}) 
for our vortex source.

\section{Numerical Calculations}

\begin{figure}
\epsfysize 7cm
\epsfbox[20 300 520 750]{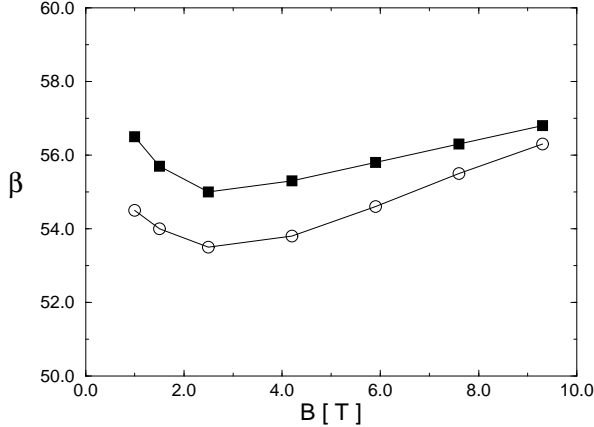}
\caption[]{Apex angle $\beta$ as a function of magnetic field $B$
at $T=0$. Circles represent the result of the calculation using
only the nonlocal corrections \cite{franz}.  
Squares correspond to the calculations considering both nonlinear
and nonlocal corrections.}
\label{fig2}
\end{figure}

The numerical calculations are performed by employing Eq.(\ref{bk}) to 
calculate $B_{\vec k}$ iteratively. At each step $G_{\rm nl}$ is calculated
numerically using FFT. In our numerical calculation, we use
$\lambda_0=1400$\AA,  $\kappa=\lambda_0/\xi_0=68$ and $\gamma=2$.
Unlike the Ginzburg-Landau free energy, the London free 
energy can not completely
determine the vortex lattice by a simple minimization. Instead, 
one has to impose a set of source terms located at the position of 
the vortices. The functional form of the source terms does not come from
the London theory and requires more fundamental treatment.
The free energy must then be minimized with respect to the positions of
the vortices (determined by the source terms).
In general, a 2D lattice can be determined by four parameters.
However, a centered rectangular lattice is energetically more favorable 
than an oblique lattice. On the other hand,
the vortex lattice spacing is fixed
by the average magnetic field $\bar B$ ($\bar B \approx H$ 
away from $H_{c1}$). Thus we are left with two variational parameters i.e.
the lattice orientation with respect to $a$ and $b$ directions and the 
apex angle $\beta$ - the angle between the two basic vectors of the lattice. 
We therefore
find the vortex lattice geometry by minimizing $\cal F$ in Eq.(\ref{fe})
with respect to the apex angle $\beta$ for different orientations of
the lattice. At $T=0$, the stable orientation of the lattice 
is aligned with the $a$ and $b$ axes. Fig.\ref{fig2}
shows the results of our numerical calculations for $\beta$ as a function of 
magnetic field. The upper curve (squares)  is the result of the combined
calculation considering both nonlocal and nonlinear
corrections i.e. using Eq.(\ref{bk}) 
and Eq.(\ref{fe}). The lower curve (circles) on the other hand, corresponds 
to taking into account only the nonlocal corrections, i.e.
neglecting $\cal F_{\rm nl}$ in Eq.(\ref{fe}) and $G_{\rm nl}$ in Eq.(\ref{bk}).
As it is clear from Fig.\ref{fig2}, the difference between the two 
cases is small and about one or two degrees. Therefore, the phase diagram
given in Ref.\cite{franz} retains its validity qualitatively even after 
adding the nonlinear correction to the London free energy.  

\begin{figure}
\epsfysize 7cm
\epsfbox[20 300 520 750]{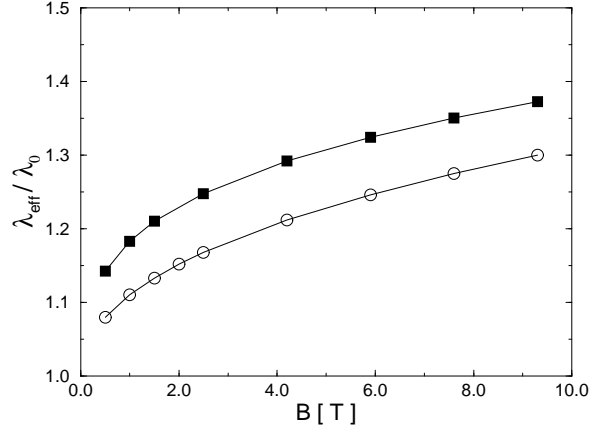}
\caption[]{The effective penetration depth as a function of the
magnetic field. Circles represent the result of the calculations
considering only the nonlocal effects  whereas squares are the result
of combined calculation considering both nonlocal and nonlinear 
effects.}
\label{fig3}
\end{figure}

We also calculate the effective penetration depth $\lambda_{\rm eff}$
for different magnetic fields in almost the same way as it is calculated
from $\mu$SR data \cite{sonier-thesis}. In these experiments, the 
$\mu$SR precession signal obtained from the experiment is fit to a signal
obtained by Fourier transforming the magnetic field distribution
calculated on a hexagonal vortex lattice with the same average magnetic
field using the ordinary London model with some cutoff function. 
The $\lambda$ which provides the best fit to the data is considered as
$\lambda_{\rm eff}$. Here, we calculate $\lambda_{\rm eff}$ in a different
way (but similar in spirit) using the fact that in the ordinary London model,
for a hexagonal lattice and for a large enough field,
\begin{eqnarray}
\overline{(B-\bar B)^2}\equiv \overline{\Delta B^2}&=&\bar B^2 
\sum_{\vec k \neq 0}{e^{-\xi^2 k^2} \over (1+\lambda^2 k^2)^2}\nonumber\\ 
&\simeq& \lambda^{-4}\bar B^2 \sum_{\vec k \neq 0}
{e^{-\xi^2 k^2} \over k^4}\propto \lambda^{-4}
\end{eqnarray}
Therefore, associating all the field dependence of $\overline{\Delta B^2}$ 
in our calculation with the field dependence of an effective penetration depth 
$\lambda_{\rm eff}$, we can define $\lambda_{\rm eff}$ by
\beq
{\lambda_{\rm eff} \over \lambda_0} = \left({\overline{\Delta B_0^2} \over
\overline{\Delta B^2}}\right)^{1 \over 4}
\label{lmd}
\eeq
where $\overline{\Delta B_0^2}$ is the mean squared value of the magnetic
field $B_0(\vec r)-\bar B$ obtained by applying  the ordinary 
London model on a hexagonal
lattice with the same average field $\bar B$ and with the penetration 
depth $\lambda_0$. 

Fig.\ref{fig3} shows the result of our numerical calculation for
$\lambda_{\rm eff}$ using Eq.(\ref{lmd}). The lower curve corresponds to
the calculations including only the nonlocal correction. The upper curve  
corresponds to the result of the calculations using 
both nonlinear and nonlocal terms. 
The effect of the nonlinear term to the field dependence of $\lambda_{\rm eff}$ 
is almost nothing but an overall shift.
Fig.\ref{fig4} exhibits magnetic field distribution $n(B)$ defined by
\beq
n(B^\prime)={1 \over A}\int d^2r \delta[B^\prime - B({\vec r})]
\eeq
at the average magnetic field $\bar B=5.9T$, 
with $A$ being the area of a unit cell.
The solid line in Fig.\ref{fig4} represents the
magnetic field distribution calculated from the nonlinear-nonlocal London
equation. This line-shape is then compared with another line-shape 
(dashed line) obtained
from an ordinary London calculation but with a larger value of $\lambda_0$.
$\mu$SR experiments also produce the same kind of line-shape but 
with some additional broadening due to  lattice disorder,
interaction of muons with nuclear dipolar fields and finite lifetime of muons.
The resolution of the magnetic field as a result of this broadening is 
$\delta B \sim 10^{-3} T$. 
The difference between the solid line and dashed line in Fig.\ref{fig4}
as well as the double peek feature of the solid line is therefore
not observable by $\mu$SR experiments because of these broadening
effects. Thus as far as these line-shapes are concerned, 
it is difficult to distinguish
a nonlinear-nonlocal effect from a simple shift in the magnetic
penetration depth in the ordinary London model. 
Fig.\ref{fig5} compares the effect of
including both nonlinear and nonlocal corrections to the London equation
with the effect of including only the nonlocal term. Comparing
the two line-shapes, it is apparent that the effect 
of the nonlinear term is small compared to the nonlocal term as was 
emphasized before.

\section{Discussion}

\begin{figure}[t]
\epsfysize 7cm
\epsfbox[20 300 520 749]{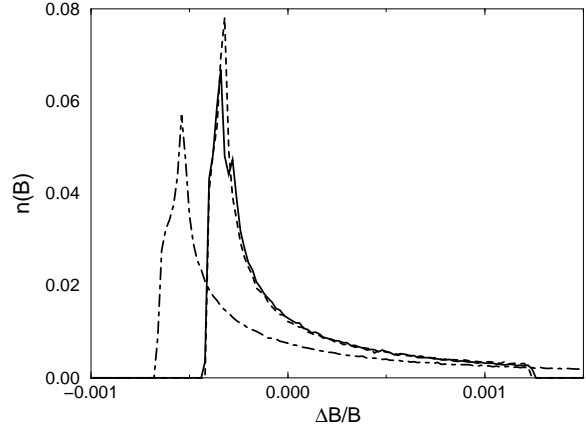}
\caption[]{Solid line:
Magnetic field distribution obtained from the nonlinear-nonlocal London
equation with $\bar B=5.9T$ and $\lambda_0=1400$ \AA. 
Dot-dashed line: Field distribution obtained from an ordinary
London equation on a hexagonal lattice with the same $\bar B$ and $\lambda_0$.
Dashed line: Result of the same ordinary London calculation
but with $\lambda_0=1850$ \AA.}
\label{fig4}
\end{figure}

The ordinary London equation is not adequate to describe all the
different properties of a vortex lattice in high $T_c$ superconductors
especially the properties resulting from the presence of the 
superconducting gap nodes or other
anisotropies on the Fermi surface. However, as we discussed in this paper 
and also in our previous publications \cite{affleck,franz}, a generalized
London model with appropriate higher order corrections which
take into account these anisotropic effects, can still
provide a fairly simple way to calculate different properties 
of a vortex lattice.

In Ref.\cite{affleck}, we established a generalized London equation 
which could describe the structure of the vortex lattice below $T_c$
down to intermediate temperatures.  Our results in Ref.\cite{affleck}
were in agreement with the Ginzburg-Landau calculations
\cite{berlinsky}. At low temperatures however, the suggested generalized
London model ceases to be valid because of the nonanalyticities resulting 
from the presence of the nodes in the superconducting gap.
In the present paper and also in Ref.\cite{franz},
we generalized our London approach to describe the nonanalytic
and also nonlinear and nonlocal nature of a d-wave vortex lattice
at low temperatures.  Our numerical calculations indicate that 
at both high temperature and low temperature regimes, the nonlocal
parts of the free energy plays the dominant role in determining the 
vortex lattice properties. 

\begin{figure}[t]
\epsfysize 7cm
\epsfbox[20 300 520 749]{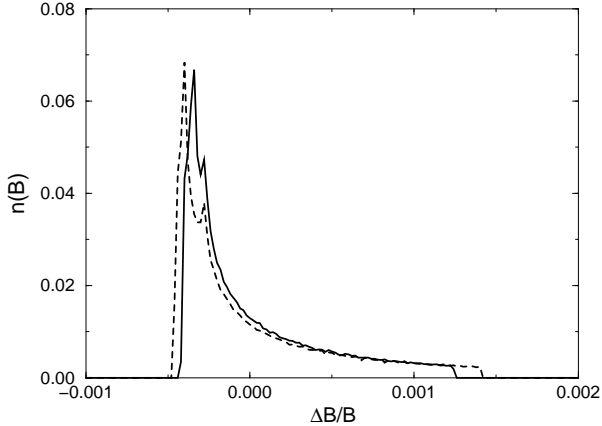}
\caption[]{Solid line: Magnetic field distribution obtained from the
nonlinear-nonlocal London equation. Dashed line: Magnetic field
distribution obtained from a London equation including only the 
additional nonlocal term using the same parameters.}
\label{fig5}
\end{figure}

The equilibrium vortex lattice geometry exhibits novel field
and temperature dependence owing to the fourfold  
anisotropic effects expressed by the corrections to the London 
equation (Fig.\ref{fig2}, also see Ref.\cite{affleck,franz}).
The numerical calculation of  the lattice geometry is rather 
insensitive to the details of the vortex cores\cite{note}. The reason is that  
the details of the magnetic field inside the vortex cores
mainly affect the self-energies of the vortices. 
In the magnetic fields far below $H_{c2}$, the vortex 
lattice geometry is mostly determined by the magnetic interaction 
energy between vortices which is insensitive to the precise shape of the core.
Therefore, our replacement  of the vortex core by a  Gaussian source 
term should be adequate for the vortex lattice structure calculations.

The effective penetration depth $\lambda_{\rm eff}$ also exhibits field
dependence at low temperatures as it is illustrated in Fig.\ref{fig3}.
Some important points need to
be emphasized here:  

(i) The field dependence of $\lambda_{\rm eff}$
is not linear. The variation of $\lambda_{\rm eff}$ with the magnetic field is
faster at lower fields. $\mu$SR data is only available
for a limited range of the magnetic field and the uncertainty of the
experimental data is noticeable. Thus, it is hard to judge from the $\mu$SR
data about the linearity of the field dependence although a negative 
curvature comparable to our result in Fig.\ref{fig3} is evident from the
$\mu$SR data for a detwinned YBCO single crystal in Ref.\cite{sonier2}. 
At low magnetic fields
where the $\mu$SR data is available, the relative variation of the
effective penetration depth in our calculation is about 7\% for an
increase of 1T in the magnetic field which is close to 7.3\% variation
obtained from $\mu$SR data for an optimally doped and 9.5\% variation
for a detwinned underdoped YBCO single crystal
using the same cutoff function as Eq.(\ref{cutoff}) for fitting calculations
\cite{sonier-thesis,sonier2}.
\begin{figure}[t]
\epsfysize 7cm
\epsfbox[20 300 540 750]{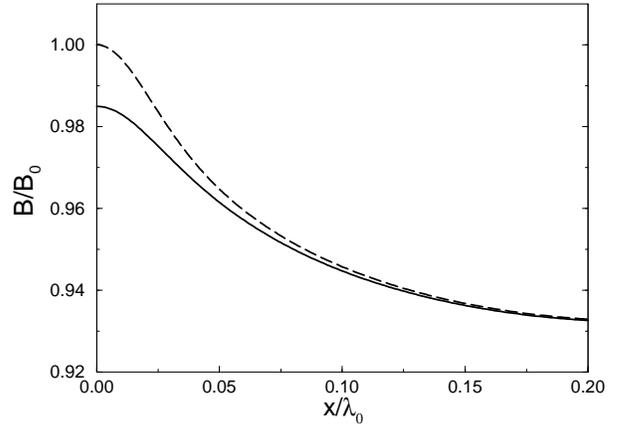}
\caption[]{Magnetic field as a function of the distance from the center of 
the vortex ($a$-direction) for an isolated vortex. The solid line corresponds
to the nonlocal London equation whereas the dashed line represents an
ordinary London calculation with the same value of $\lambda_0$. 
$B_0=B(r=0)$ for the ordinary London vortex.}
\label{fig6}
\end{figure}

(ii) More importantly, 
this field dependence of $\lambda_{\rm eff}$ has a predominantly nonlocal
origin rather than a nonlinear one, contrary to what is generally believed.
The contribution of the nonlinear term to the total (minimized) 
free energy is almost
one order of magnitude smaller than the nonlocal term. What is more important
however, is the field dependence and $\beta$ dependence of these terms
not their orders of magnitude at
fixed $\beta$ and B. As we mentioned earlier, we consider $\beta$ as a variational
parameter which has to  be fixed by minimizing the London free energy.
As can be inferred from Fig.\ref{fig2} and Fig.\ref{fig3},
the field dependence and $\beta$ dependence of the nonlinear 
term in the free energy is also smaller than the nonlocal term. 
It is worth noting that in the Meissner state, a linear nonlocal 
term can never produce field dependence in the penetration depth 
(as it is usually defined in that state) and
therefore a nonlinear term is necessary for such an effect\cite{ys}.
In the vortex state however, a nonlocal term can result in a field dependent
effective penetration depth. To understand this, let's neglect the
nonlinear term and assume a linear but nonlocal London equation.
In that case the total magnetic field will be the superposition of the 
fields around
individual vortices. The magnetic field around an isolated vortex is given by
\beq
B(\vec r)=\Phi_0\int {d^2k \over (2\pi)^2} {F(k)e^{i\vec k . \vec r} \over 
1+{\cal L}_{ij}(\vec k)k_ik_j}
\label{sglvx}
\eeq
where $\Phi_0$ is the flux quantum and $F(k)$ is the cutoff function 
resulting from the source term. ${\cal L}_{ij}(\vec k)$ is defined
in Eq.(\ref{L}). For small values of 
$k$, ${\cal L}_{ij}(\vec k) \approx \lambda_o^2 \delta_{ij}$. Since small
$k$ corresponds to large $r$, one 
expects an isotropic field, similar to the local London case, far away 
from the vortex core. For large values of $k$ however, 
${\cal L}_{ij}(\vec k)$ has
strong $k$-dependence with four-fold anisotropy. Thus the closer to 
the vortex core, the more deviation form an isotropic ordinary London single 
vortex is expected, as is clearly shown in Fig.\ref{fig6}.
At low magnetic fields, the vortices are far apart and
their magnetic fields overlap in regions far away from their cores. The 
properties of the vortex lattice should then be similar to the ordinary 
london hexagonal lattice. As the magnetic field is increased, the vortices
come closer to each other. Although the profile of the magnetic field 
around each vortex remains unchanged (the nonlinear term is neglected), 
the overlap regions will be closer to
the vortex cores and will be more affected by the nonlocal term. Therefore,
it is conceivable that at large magnetic fields, the vortex lattice 
properties such as the magnetic field distribution will be affected by 
the nonlocal term in the London equation. The magnetic field near the 
vortex core is reduces by the nonlocal term as is clear in 
Fig.\ref{fig6}. This is because ${\cal L}_{ij}(\vec k)-\lambda_0^2\delta_{ij}$ 
is a positive definite tensor for all $k$ and therefore the denominator
of the integrand in Eq.(\ref{sglvx}) is always larger 
than the corresponding ordinary London one.
As a result, the magnitude of $\overline{\Delta B^2}$ is smaller for 
the nonlocal case and therefore $\lambda_{\rm eff}$ defined by
Eq.(\ref{lmd}) tends to be larger. This explains why $\lambda_{\rm eff}/\lambda_0$
is always greater than one in Fig.\ref{fig3}. Fig.\ref{fig4} exhibits the
resemblance between a change in the magnetic field distribution due 
to the nonlocal term and due to 
a shift in the ordinary London penetration depth.
The slight difference between the solid and dashed lines in Fig.\ref{fig4} would
be unobservable in $\mu$SR experiments as a result of the broadening effects.
Since no field dependence due to the nonlocal term is expected
in the Meissner state, 
the $\lambda_{\rm eff}$ as defined here and also in $\mu$SR experiments
is expected to be conceptually different
from what is usually defined as penetration depth in the Meissner state,
although they are closely related.

(iii) The magnitude and field dependence of $\lambda_{\rm eff}$ is not
so sensitive to the apex angle $\beta$. In other words, a few degrees 
change in the variational parameter $\beta$ does not modify the 
magnetic field distribution 
as much as a variation in the average magnetic field does.  

(iv) Calculation of $\lambda_{\rm eff}$ is rather sensitive to the
form of the vortex source term. The importance of the source term in the
calculation of $\overline{\Delta B^2}$ has already been 
emphasized by Yaouanc {\em et al.}\cite{yaouanc}. 
In Ref.\cite{sonier-thesis}, $\lambda_{\rm eff}$ is obtained by fitting to 
the $\mu$SR data using both a Gaussian cutoff (Eq.(\ref{cutoff})) and also the
cutoff function proposed by Hao {\em et al.}\cite{hao,yaouanc}. The 
difference between the two cases is significant and about 30\% for the
magnitude of $\lambda_{\rm eff}$ and even more (for detwinned sample)
for the relative variation with respect to
the magnetic field. This can explain
the importance of the source term in calculations of the effective 
penetration depth.

$\mu$SR experiments \cite{sonier-thesis,sonier3} on NbSe$_2$, which 
is believed to be a conventional
superconductor, also show a field dependence in the effective penetration depth
although it is much weaker than what is observed  in high $T_c$ 
compounds. Since there
is no node in the superconducting gap of these materials, the theory
presented in this article can not explain this field dependence.  
However, since the size of the vortex core in these materials is large
and comparable to the vortex lattice spacing for the magnetic fields of
experimental interest, it is conceivable that a significant effect can come
from the cores as it is pointed out in Ref.\cite{yaouanc}. 
Thus, a more careful consideration
of the vortex core might be necessary in order to have a better quantitative
explanation of the experimental results \cite{amin}.
 
\section{Conclusion}

We investigated the effect of the superconducting gap nodes on the 
vortex lattice properties at very low temperatures by a generalized
London approach with higher order corrections to the free energy.
We found that nonlocal effects, arising from the diverging coherence length 
near the gap node, are predominantly responsible for the unusual behavior of 
the vortex lattice geometry and the effective penetration depth. 
The nonlinear effects  associated with the shift of the quasiparticle 
excitation spectrum play only a secondary role, resulting in small (but not
negligible) corrections at low $T$.
Contrary to the common belief, the effective penetration depth, as 
defined in a $\mu$SR 
experiment, is not a linear function of the magnetic
field and is mainly affected by the nonlocal effects. This is in a marked
contrast to the nonlinear Meissner effect\cite{ys} in a $d$-wave superconductor
where the correction to $\lambda_{\rm eff}$ arises strictly from the
nonlinear term in the London free energy and is linear in the field.

\section{Acknowledgment}

We would like to thank P. Stamp, W. Hardy, R. Kiefl, J. Sonier, I. Herbut, 
M. Sigrist, R. Heeb 
and Z. Te\v{s}anovi\'c for useful discussions, 
M. Kohmoto and P. Muzikar 
for correspondence and M. Nekovee for numerical advise.
This works was supported by NSERC, the CIAR and NSF
grant DMR-9415549 (M.F).


\begin{references}
\bibitem{note1} There is a factor of -1/2 difference from our previous notation
\cite{affleck}. Here, we follow \cite{ys} by setting the electron mass equal 
to one.
\bibitem{note} 
The calculations presented here and in Ref.\cite{affleck,franz} 
have used isotropic core
cut-off functions.  For d-wave superconductors the actual vortex cores
will not necessarily be isotropic \cite{Maki}.  We are presently
investigating the effect of core anisotropy on vortex lattice geometry
using the modified London free energy approach.
\end{references}
\end{document}